\documentclass[conference]{IEEEtran}
\IEEEoverridecommandlockouts
\usepackage{url}
\usepackage{balance}
\usepackage{booktabs}
\usepackage{multirow}
\usepackage{tcolorbox}

\usepackage{cite}
\usepackage{amsmath,amssymb,amsfonts}
\usepackage{algorithmic}
\usepackage{graphicx}
\usepackage{textcomp}
\usepackage{xcolor}
\def\BibTeX{{\rm B\kern-.05em{\sc i\kern-.025em b}\kern-.08em
    T\kern-.1667em\lower.7ex\hbox{E}\kern-.125emX}}

\def\mybar#1{
  {\color{gray}\rule{#1cm}{9pt}}}

\newcommand{\ShLog}{Log4JShell}
\newcommand{\ShLogRe}{Log4JShell-related}

\begin{document}

\title{Drop it All or Pick it Up? How Developers Responded  to the Log4JShell Vulnerability}



\author{\IEEEauthorblockN{Vittunyuta Maeprasart, Ali Ouni*, Raula Gaikovina Kula}
\IEEEauthorblockA{
\textit{Nara Institute of Science and Technology (NAIST)}\\
\textit{École de technologie supérieure, Montreal*}\\
\{vittunyuta@gmail.com , raula-k\}@is.naist.jp}
ouniaali@gmail.com*\\
}

\maketitle

\begin{abstract}
Although using third-party libraries has become prevalent in contemporary software development, developers often struggle to update their dependencies.
Prior works acknowledge that due to the migration effort, priority and other issues cause lags in the migration process.
The common assumption is that developers should drop all other activities and prioritize fixing the vulnerability.
Our objective is to understand developer behavior when facing high-risk vulnerabilities in their code.
We explore the prolific, and possibly one of the cases of the \texttt{\ShLog}, a vulnerability that has the highest severity rating ever, which received widespread media attention. Using a mixed-method approach, we analyze 219 GitHub Pull Requests (PR) and 354 issues belonging to 53  Maven projects affected by the \texttt{\ShLog} vulnerability.
Our study confirms that developers show a quick response taking from 5 to 6 days.
However, instead of dropping everything, surprisingly developer activities tend to increase for all pending issues and PRs. 
Developer discussions involved either giving information (29.3\%) and seeking information (20.6\%), which is missing in existing support tools.
Leveraging this possibly-one of a kind event, insights opens up a new line of research, causing us to rethink best practices and what developers need in order to efficiently fix vulnerabilities.
\end{abstract}

\section{Introduction}
\label{sec:intro}
Developers are increasingly becoming aware of the ever-growing threat of security vulnerabilities in their third-party dependencies \cite{Web:octverse}. 
In this age of social media, and information alerts, developers can be notified quickly of when their applications are at risk.
For example, security advisories allow maintainers and users to be notified of any potential malicious exploits.
Much research work and recent technological advancements has been invested in detecting (e.g., Eclipse Steady \cite{Ponta:icsme2018,PontaEMSE2020}), and fixing (e.g., Dependabot \cite{Dependabot:online},
Greenkeeper.io \cite{Greenkeeper:online},
Renovate \cite{renovate:online}).
These mechanisms allow developers to become aware of affected components so they can quickly mitigate the risk.

Although such awareness have improved over recent times, most studies still report lags in the update in response to third-party components.
For instance, a plethora of work \cite{Kula:2017,Bodin:EMSE2021,Alfadel:SANER2021,Decan:2018,Cox-ICSE2015,Li:2017,Linares:2017,ZeroualiArxiv2021,Piantadosi:2019,RuohonenIWESEP2018} report lags in the update of security updates and has generally been the case for any dependency update in the ecosystem.
Developers often cite roles, being undermanned, and the migration effort needed to mitigate the risk \cite{kula2018empirical,Bogart:2015,BogartFSE2016,Bavota:2015,Hora:2015,alrubaye2019use}.

Our aim is to analyze and understand developers behavior in a case where developers were pressured to mitigate a vulnerability, due to its high severity and profiling nature.
In particular, we study the very rare and prolific case of \ShLog~ vulnerability \cite{web:SecurityLog4Shell2021}, that affects any application that uses the most popular Java logging library Log4j.
Assigned the unique security vulnerability id CVE-2021–44228 \cite{web:CVE202144228},  the \ShLog~vulnerability allows an attacker to inject in the requests that get parsed and executed by the vulnerable application. The vulnerability was assigned the highest severity rating as its exploitation is not only likely to results in root-level compromise of servers or infrastructure devices, but also straightforward where the attackers do not require any special authentication credentials or knowledge.
Resources available on social media streams such as Twitter, Reddit, and YouTube raised the profile and awareness of this epic vulnerability.
The severity was raised further by the fact that mitigation options are limited when it comes to Java logging libraries.
This vulnerability provides an opportunity to better understand how developers respond and what issues are faced during mitigation. Our study was driven by these two following research questions:
\begin{itemize}
\item \textbf{RQ1: \textbf{How do developers respond to disclosure of a vulnerability in terms of time and activity?}}
\textit{Motivation:}
Recent work \cite{Bodin:EMSE2021} show a lag in security updates. We confirm whether this also applies to a critical severity vulnerability, and its effect on other development activities.
\item\textbf{RQ2: \textbf{What is the content of developers’ discussions while resolving a vulnerability?}}
\textit{Motivation:}
Our motivation is to understand whether there are some insights into the information needs that developers require to fix the threat.
\end{itemize}

An unexpected observation made when analyzing RQ1 was that instead of developers dropping everything to fix the vulnerability, developers tend to speed up their development activities.
Complementary, the results of RQ2 show that developers spend efforts discussion the vulnerability, either providing or seeking the needed information.  

\section{Background}
This section provides background information to our study.

\textbf{ What is \texttt{\ShLog}?} 
Early in December 2021, numerous cybersecurity researchers and media outlet  \cite{web:ShiftLog4Shell2021, web:WhatILearn2021,web:GitHubCVE2021} began sounding the alarm about a vulnerability that was later classified as critical zero-day exploit in the Java logging library Apache Log4j. Assigned a unique id \texttt{CVE-2021-44228}, the \ShLog~ vulnerability affected the Open Source Maven library Apache log4J \cite{Log4jApache:online} 
with the official description from the security advisory description where:
\begin{quote}
    \textit{``...An attacker who can control log messages or log message parameters can execute arbitrary code loaded from servers when message lookup substitution is enabled...'' \cite{web:CVE202144228} }
\end{quote}

\begin{table*}[]
\centering
\caption{Statistics of the collected Maven libraries}
\label{tab:dataset_statistic}
\begin{tabular}{@{}lr@{}}
\toprule 
\multicolumn{2}{c}{\textbf{Data Processing Information}} \\ \midrule
snapshot period &  from 1st-Dec-2021  \\
 &  to 27th-Dec-2021 \\
\# dependent on log4j library & 367\\ \hline
has GitHub repo &  227\\ 
\# contain 1 PR & 89 \\
\# contain 1 Issue & 71 \\ 
\# dependents (mean, median, min, max, sd) & (18, 0, 0, 506, 65) \\
\# dependencies (mean, median, min, max, sd) & (588, 1, 0, 37645, 3328) \\
\# star count (mean, median, min, max, sd) & (1176, 17, 0, 46234, 4954) \\ 
\# contain \ShLogRe~PR/issue & 53 \\
\hline
\# Pull Requests (RQ1) & 1827 \\
\hspace{1em} - \# \ShLog~PRs & 219 (11.99\%) \\ 
\hspace{1em} - \# Other PRs  & 1608 (88.01\%) \\
\# Non-bot PRs & 1396	(76.41\%) \\ 
\# Bot PRs  & 431 (23.59\%) \\ \hline
\# Issues & 2631 \\
\hspace{1em} - \# \ShLog~issues & 354 (13.45\%) \\ 
\hspace{1em} - \# \ShLog~ comments (RQ2) & 208 \\ 

\bottomrule
\end{tabular}
\end{table*}

\textbf{{How is a client application compromised?} } 
The vulnerability was introduced with the new Java Naming and Directory Interface (JNDI) lookup feature that allows any inputs to be parsed and interpreted by the client application no matter where it originates. JNDI is an API that allows the Java application to perform searches on objects based on their names \cite{JNDI:online}.
It has been reported that some hackers exploit this vulnerability to utilize the capabilities of the victims' devices; uses include cryptocurrency mining, creating botnets, sending spam, establishing backdoors and other illegal activities such as ransomware attacks \cite{web:SecurityLog4Shell2021,web:WhatILearn2021}.

\textbf{{What client applications are targeted?} } 
The attacker sets up a rogue server, then can create an exploit payload class, and stores it as an object such as ``\texttt{Log4JPayload.class}'' to get referenced later. 
Then, the attacker inserts the crafted JNDI injection to any requests that are likely to be logged, such as the request paths, HTTP headers, Filenames, Document/Images EXIF and so on.
An attacker can then leverage the information to choose the appropriate attack vector to compromise the targeted application.
It can affect web applications, databases, email servers, routers, endpoint agents, mobile apps, IoT devices — you name it (if it runs Java, it could be vulnerable).


\textbf{{How to mitigate the vulnerability?}} 
We found four reported methods to mitigate the vulnerabiity: 1)
Upgrade to the latest version of Log4J — v2.17.0, 2) Disable lookups within messages\footnote{Set up \texttt{log4j2.formatMsgNoLookups=true}}, 3) Remove the JndiLookup class from the classpath\footnote{zip -q -d log4j-core-*.jar org/apache/logging/log4j/core/lookup/JndiLookup.class}, or 4) Apply firewall rules to limit communications to only a few allowable hosts, not with everyone.
Out of all of the methods, upgrading to a safer release is the most efficient in our opinion.

\section{Effect on  the Maven Ecosystem}
In this section, we present our empirical study design that includes the data sources, collection and processing.
We choose the Maven set of libraries as our case study since they are highly relied upon by Java applications, and also are active Open Source projects that are important to the Java ecosystem.

\textbf{Data sources:}
There are two data sources to mine the dataset. 
The first is to gain the information related to Maven libraries where we use the libraries.io \cite{Librariesio:online}.
Then, we use GitHub graphql \cite{GraphQL:online}
to capture PR and issues information. 

\textbf{Identifying Vulnerable Projects and their Fixes:}
As shown in Table \ref{tab:dataset_statistic}, we first queried \textit{libraries.io} to find 367 Maven projects that depended on log4j-core. 
We then identified 227 Maven projects that had their code hosted on GitHub.
Using the GitHub API, we then queried for active projects during the time period of interest (1st Dec to 27th Dec 2021).
We chose this time period since, this was a critical period from which the CVE-2021-44228 advisory was released.
In the end, we collected the Pull Requests that belong to these 89 affected Maven projects.

To identify the PRs and issues related to the vulnerability, we then conducted a simple regular expression search in both the title and description, using the following search string "log4j".
From this search, we identified 53 repositories having 219 PRs and 354 issues.

Another interesting observation we found with these repositories is that they had other pending PRs and issues that developers also had to resolve.

Furthermore, note that many projects raised multiple \ShLogRe~PRs and issues.
An appendix of the data collected is available in our replication package: \url{https://doi.org/10.5281/zenodo.6030472}.

\begin{figure}[!]
    \centering
    \includegraphics[width=\linewidth]{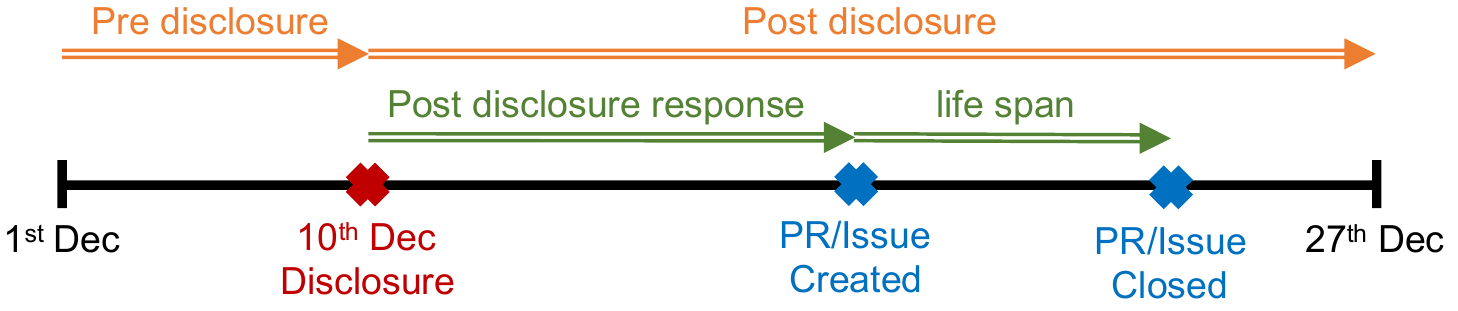}
    \caption{Terminology used to describe the timeline pre and post the disclosure of \ShLog}
    \label{fig:timeline_definition}
\end{figure}

\section{Developer Response-time And Activity (RQ1)}
To answer RQ1, we first present the following terminology to assist with our analysis, as shown in Figure \ref{fig:timeline_definition}.
\begin{itemize}
    \item \textit{Disclosure} - is the date (10th Dec-2021) when the security advisory for CVE-2021-44228 was released. 
    \item \textit{pre-disclosure} - is the time period before  disclosure release.
    \item \textit{post-disclosure} - is the time period after disclosure release.
    \item \textit{post-disclosure response} - is the time period since the disclosure to when a PR was created.
    \item \textit{PR lifespan} - is the time period from when a PR was created to when it was closed. 
\end{itemize}

\begin{table*}[]
\centering
\caption{Reponse-time statistics of \ShLogRe~ PRs}
\label{tab:log4j_time_statistic}
\scalebox{1}{
\begin{tabular}{l|rrrrr}
\toprule
\multicolumn{1}{c|}{\textbf{PR}} & \multicolumn{1}{c}{\textbf{Mean}} & \multicolumn{1}{c}{\textbf{Median}} & \multicolumn{1}{c}{\textbf{SD}} & \multicolumn{1}{c}{\textbf{Max}} \\ \hline
Post disclosure response (\#days)       & 5.31             & 4.88               & 3.39                     & 14.17       \\
PR life span (\#days)            & 1.30             & 0.50               & 1.76                      & 6.46    \\
Post disclosure response plus     & 6.35             & 5.97               & 3.31                                       & 13.91           \\   
\bottomrule
\end{tabular}}
\end{table*}

Then, we employ provide three different quantitative analysis methods as follows.

\textbf{Analysis \#1: Response to the Disclosure.}
The motivation for this first analysis is to understand the time spent before a project receives a PR to mitigate the vulnerability. 
Table \ref{tab:log4j_time_statistic} shows the summary statistics of how long it took developers to merge any responses to fix the vulnerability fix. 
We can see that it takes around four to five days on average after the disclosure before the PR is raised (i.e., median of 4.88).
However, the fix itself is quickly resolved in about 1 day on average to get accepted and merged. 
This leaves the whole process lasting five to six days.

\begin{figure}[!]
    \centering
    \includegraphics[width=\linewidth]{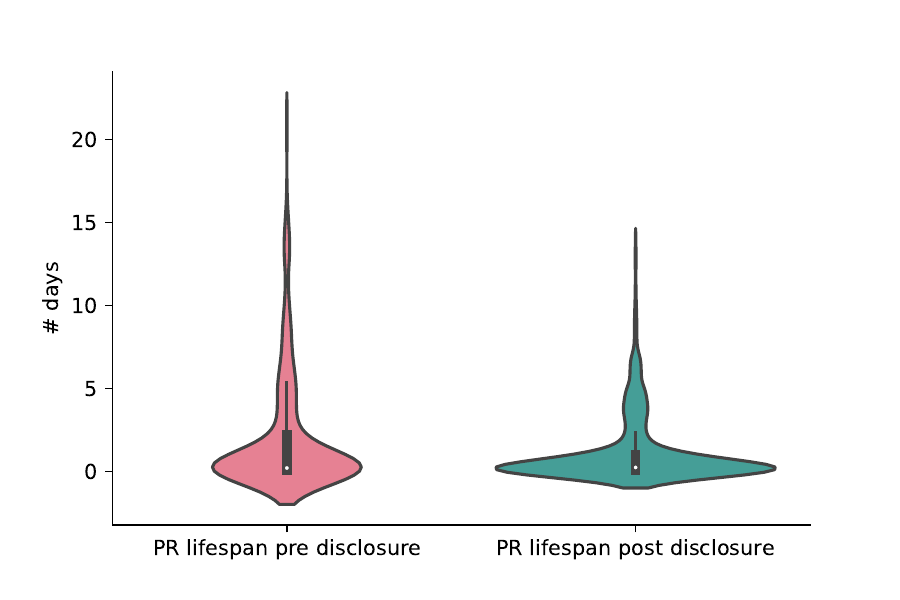}
    \caption{Comparing lifespan of PRs pre and post disclosure}
    \label{fig:rq1_2}
\vspace{-.3cm}
\end{figure}

\textbf{Analysis \#2: Effect of the Disclosure on other activities.}
The motivation for the second analysis is to understand whether or not the disclosure had any effect on PR lifespan. 
Therefore, we would like to test whether there may be an increase in developer activity, and whether PRs might be resolved quicker.
To perform this analysis, we compare the lifespan of all PRs that were created before and after the disclosure.
In the end, we identified 500 PRs created before the disclosure and 948 PRs created after the disclosure (166 \ShLogRe). 

As shown in Figure \ref{fig:rq1_2}, we find that developers seem to resolve all PRs (both \ShLogRe~and those not) quicker after the disclosure. 
To assess the statistical difference between \ShLogRe~ related PRs and other PRs, we apply the Mann-Whitney U test \cite{mann1947test}, which is a non-parametric test that is used to compare two sample means come from the same population. We also measure the effect size using Cliff’s $\delta$, a non-parametric effect size measure \cite{cliff1993dominance}.
Effect size is analyzed as follows: 
(1) |$\delta$| $<$ 0.147 as Negligible, 
(2) 0.147 $\leq$ |$\delta$| $<$ 0.33 as Small, 
(3) 0.33 $\leq$ |$\delta$| $<$ 0.474 as Medium, or 
(4) 0.474 $\leq$ |$\delta$| as Large. We use the cliffsDelta \cite{neilerns23:online}
package to analyze Cliff’s $\delta$.
Overall, we found a statistical difference between PRs created before and after the disclosure, confirming that developers had increased their development activities in general (\textit{p}-value < 0.01, Cliff’s $\delta$ is Large). 

\begin{figure}[!]
    \centering
    \includegraphics[width=\linewidth]{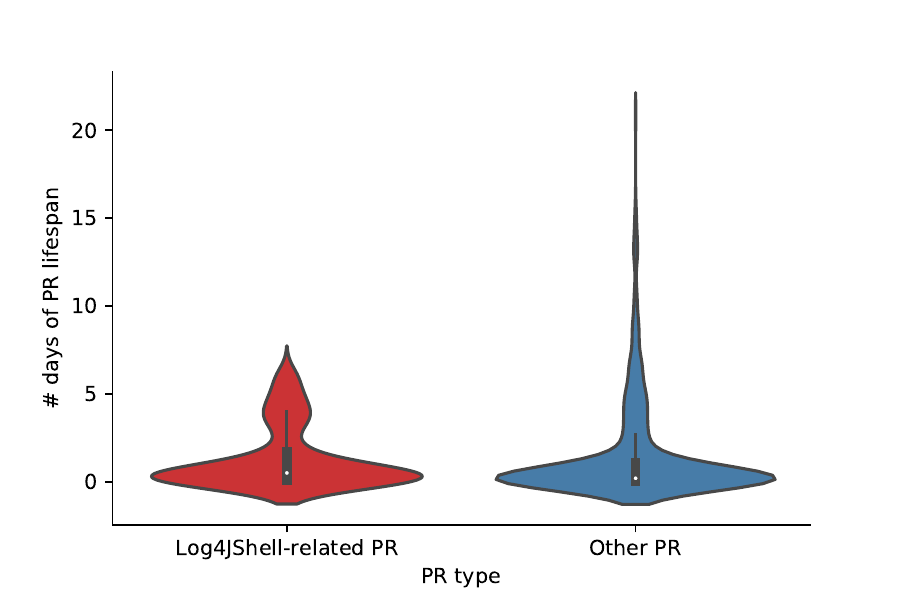}
    \caption{Comparing lifespan of \ShLogRe~against other PRs.}
    \label{fig:rq1_3}
\end{figure}

\begin{table*}[]
\centering
\caption{Comparison of PR patches between \ShLogRe~ and other PRs}
\label{tab:log4j_statistic}
\scalebox{1}{
\begin{tabular}{l|rlrlrlrlrl|lc}
\toprule
\multicolumn{1}{c|}{\multirow{2}{*}{\textbf{PR variables}}} & \multicolumn{10}{c|}{\textbf{Statistic of Log4j PR vs Other PR}} & \multicolumn{2}{c}{\textbf{Statistical value}}              \\ \cline{2-13} 
\multicolumn{1}{c|}{}                                       & \multicolumn{2}{c|}{\textbf{Mean}}                       & \multicolumn{2}{c|}{\textbf{Median}}            & \multicolumn{2}{c|}{\textbf{SD}}                         & \multicolumn{2}{c|}{\textbf{Min}}               & \multicolumn{2}{c|}{\textbf{Max}} & \textbf{p-value} & \multicolumn{1}{l}{\textbf{Cliff’s $\delta$}} \\ \hline
\# Files                                                     & \multicolumn{1}{r|}{2.54}  & \multicolumn{1}{l|}{13.68}  & \multicolumn{1}{r|}{1} & \multicolumn{1}{l|}{2} & \multicolumn{1}{r|}{6.79}   & \multicolumn{1}{l|}{92.14} & \multicolumn{1}{r|}{1} & \multicolumn{1}{l|}{0} & \multicolumn{1}{r|}{87}    & 3072 & $<$ 0.001        & medium                                        \\
\# Commits                                                   & \multicolumn{1}{r|}{1.43}  & \multicolumn{1}{l|}{2.24}   & \multicolumn{1}{r|}{1} & \multicolumn{1}{l|}{1} & \multicolumn{1}{r|}{3.65}   & \multicolumn{1}{l|}{5.41}  & \multicolumn{1}{r|}{1} & \multicolumn{1}{l|}{0} & \multicolumn{1}{r|}{54}    & 123  & $<$ 0.001        & small                                         \\
\# Added LoC                                                & \multicolumn{1}{r|}{37.75} & \multicolumn{1}{l|}{847.81} & \multicolumn{1}{r|}{1} & \multicolumn{1}{l|}{8} & \multicolumn{1}{r|}{279.44} & \multicolumn{1}{l|}{23k}   & \multicolumn{1}{r|}{0} & \multicolumn{1}{l|}{0} & \multicolumn{1}{r|}{4039}  & 935k & $<$ 0.001        & medium                                        \\
\# Deleted LoC                                              & \multicolumn{1}{r|}{28.43} & \multicolumn{1}{l|}{793.67} & \multicolumn{1}{r|}{1} & \multicolumn{1}{l|}{3} & \multicolumn{1}{r|}{254.28} & \multicolumn{1}{l|}{23k}   & \multicolumn{1}{r|}{1} & \multicolumn{1}{l|}{0} & \multicolumn{1}{r|}{3691}  & 903k & $<$ 0.001        & medium                                        \\
\# Requested reviewers                                      & \multicolumn{1}{r|}{0.16}  & \multicolumn{1}{l|}{0.18}   & \multicolumn{1}{r|}{0} & \multicolumn{1}{l|}{0} & \multicolumn{1}{r|}{0.43}   & \multicolumn{1}{l|}{0.57}  & \multicolumn{1}{r|}{0} & \multicolumn{1}{l|}{0} & \multicolumn{1}{r|}{2}     & 6    & 0.4963           & negligible                                    \\
\bottomrule
\end{tabular}
}
\end{table*}

\begin{table*}[]
\centering
\caption{ Frequency count of developer intention during vulnerability discussions}
\label{tab:rq2_taxonomy_result}
\scalebox{1}{
\begin{tabular}{lll}
\toprule
\textbf{Category}  & \textbf{Description} & \textbf{Count}\\ \midrule 
Information Giving (IG) & Share knowledge and experience with other people, or inform other people about  & 61 \mybar{1.0} \\
 &  new plans/updates (e.g. “\textit{FYI, severity on CVE-2021-45046  was increased to 9.0.}”). \cite{onlineIG} & \\
Information Seeking (IS) & Attempt to obtain information or help from other people (e.g. \textit{When are we}  & 43 \mybar{0.70} \\
& \textit{expecting this to be fixed and available on maven?}”). \cite{onlineIS2} & \\
Solution Proposal (SP) & Share possible solutions for discovered problems (e.g. “\textit{Fixed this by regular} & 38 \mybar{0.62} \\
& \textit{dependency updating process..}”). \cite{onlineSP} & \\
Meaningless (ML) & Sentences with little meaning or importance (e.g. “\textit{Thanks for the quick work.}”). \cite{web:onlineML} & 37 \mybar{0.61} \\
Aspect Evaluation (AE) & Express opinions or evaluations on a specific aspect  (e.g. “\textit{I think the robot is} & 24 \mybar{0.39} \\
& \textit{only pushing a suggested version.}”). \cite{onlineAE} & \\
Problem Discovery (PD) & Report bugs, or describe unexpected behaviors (e.g. “\textit{I tried solution suggested above}, & $\;\,$5 \mybar{0.08}\\
& \textit{but it is not working},”). \cite{onlinePD} & \\ \hline
Total Comments & &  208 \\
\bottomrule
\end{tabular}
}
\end{table*}


\textbf{Analysis \#3: \ShLogRe~PRs vs. Other PRs.}
The motivation for the final analysis is to understand whether or not \ShLogRe~PRs are resolved quicker than other PRs.
Hence, we measure the lifespan between \ShLogRe~PR and other PRs.
Our assumption is that developers may pay more attention the \ShLogRe~PRs, thus fixing them quicker.
In the end, we found 166 PRs related to the \ShLogRe, and 1,282 other PRs that were not.
Interestingly, as shown in Figure \ref{fig:rq1_3}, we find that this is not the case, with \ShLogRe~PRs having longer lifespans.
Similar to the second analysis, we use cliffs delta and the Mann-Whitney test to test our hypothesis. 
Hence, we confirm a statistical difference to show that developers spend more time closing a \ShLogRe~PR (\textit{p}-value < 0.01, Cliff’s $\delta$ is Large). 

Additionally, as shown in Table \ref{tab:log4j_statistic}, we find the \ShLogRe~PRs are smaller in size.
We report that the number of files, commits, added and deleted lines of code are all significantly smaller that other submitted PRs (p-value significant and  Cliff’s $\delta$ small to medium).

\begin{tcolorbox}
\textbf{Summary}: We find the developers response to \ShLogRe~ vulnerability to take a 5 to 6 day turnaround.
We find the that the vulnerability disclosure impacts with all PRs being resolved quicker.
Finally, we find that although the fix is relatively small, developers take more time to merge \ShLogRe~PRs when compared to other PRs.
\end{tcolorbox}

\section{Developer Discussions (RQ2)}
To answer RQ2, we take a qualitative approach to have a deeper understanding of what developers discuss when fixing \ShLog.
We adopt the taxonomy from \cite{QiaoTSE2020}, and perform multiple iterations on the dataset.
In the end, we ended up with the 208 comments.
We performed a manual analysis between three of the four authors, using a round table style to discuss and classify the comments. 
First, all three authors sat together to discuss and classify the first 30 comments. 
After that the second author then classified the rest all samples.
This was later double checked by the first author, raising any comments that they did not agree with.
Finally, in the third round, the third author then proceeded to resolve any disagreements.
The first two authors are post graduate students, while the third author is an assistant professor with over eight years of experience.

Table \ref{tab:rq2_taxonomy_result} shows that most comments raised by developers while solving was to provide information (61 instances).
Examples include FYIs, such awareness to the team that the vulnerability severity was raised \cite{onlineIG}.
Information seeking was the second most frequent (43 instances), with examples such as requests other team members (spring officials) to consider rebuilding a version based on the fix \cite{onlineIS}.
Other responses included solution proposals (38 instances), meaningless (37 instances), aspect evaluation (24 instances).
Interestingly, problem discovery was the least discussed comment (5 instances), suggesting that developers did not report as many bugs or unexpected behaviors.

\begin{tcolorbox}
\textbf{Summary}:
We find a combined 61.2\% of information needs of developers are to either give (61 out of 208) or seek information (43 out of 208) how to mitigate the vulnerability.
Developers did not report bugs, or unexpected behaviors (5 out of 208).
\end{tcolorbox}

\section{Limitations}
Since we only focus on Log4JShell, we acknowledge that generalization is a key threat. 
Furthermore, our time-frame for the data collection may cause bias to the results.
However, as in initial concept, we are confident our preliminary result can be strengthened with a complete study. 

\section{Research Outlook}
We highlight two takeaways and the research outlook.
\paragraph{Rethinking How Developers Fix Severe Threats}
Our key takeaway message is that when it comes to a critical vulnerability, instead of dropping everything, developers tend to increase their activities, resolving all pending PRs and issues before attending to the fix.
We make the argument that it is equally important to assist developers with resolving pending issues, as this affects how quick, and the resources needed to seemingly fix a single vulnerability.
For instance, it was reported on social media that the Log4J support team would have welcomed assistance \textit{"...Log4j maintainers have been working sleeplessly on mitigation measures; fixes, docs, CVE, replies to inquiries, etc...Yet nothing is stopping people to bash us, for work we aren't paid for, for a feature we all dislike yet needed to keep due to backward compatibility concerns."} 
\cite{web:Twitter2021}
We argue that research into understanding the prerequisites for fixing a vulnerability is needed.
A concrete example is tool support to inform developers of pending issues and PRs that may or may not affect a library.
This change in mindset promotes best practices to maintaining third-party interfaces to the code (Application Programming Interfaces).
How this mindset is perceived by developers and its barriers is deemed as future work.

\paragraph{Recommending Information Needs to Fix a Severe Threat}
Our results motivate the need for developers to readily access information on the vulnerability. 
Although these are existing global information hubs like the vulnerability database, and generic blogs.
Our immediate research direction is to use crowd-sourcing to search and recommend information across different projects that are resolving the same vulnerability.

\section*{Acknowledgement}
This work is supported by the Japanese Society for the Promotion of Science (JSPS) KAKENHI Grant Number JP20H05706.
\bibliographystyle{ieeetr}
\bibliography{sample-base}

\end{document}